# 3-D Tracking and Visualization of Hundreds of Pt-Co Fuel Cell Nanocatalysts During Electrochemical Aging

*Yingchao Yu,[&,†] Huolin L. Xin,[&,‡] Robert M. Hovden,[§] Deli Wang,[†] Eric D. Rus,[†] Julia Mundy,[§]*

*David A. Muller,\*[§] Héctor D. Abruña,\*[,†]*

[†]Department of Chemistry and Chemical Biology, Cornell University, Ithaca, NY 14853, USA

[‡]Department of Physics, Cornell University, Ithaca, NY 14853, USA

[§] School of Applied and Engineering Physics & Kavli Institute, Cornell University, Ithaca, NY 14853, USA

[&] These authors contributed equally to this work

\* Email: dm24@cornell.edu, hda1@cornell.edu

**ABSTRACT**

We present an electron tomography method that allows for the identification of hundreds of electrocatalyst nanoparticles with one-to-one correspondence before and after electrochemical aging. This method allows us to track, in three-dimensions (3-D), the trajectories and morphologies of each Pt-Co nanocatalyst on a fuel cell carbon support. The use of atomic-scale electron energy loss spectroscopic imaging enables the correlation of performance degradation of the catalyst with changes in particle/inter-particle morphologies, particle-support interactions and the near-surface chemical composition. We found that, aging of the catalysts under normal fuel cell operating conditions (potential scans from +0.6 V to +1.0 V for 30,000 cycles) gives rise to coarsening of the nanoparticles, mainly through coalescence, which in turn leads to the loss of performance. The observed coalescence events were found to be the result of nanoparticle migration on the carbon support during potential cycling. This method provides detailed insights into how nanocatalyst degradation occurs in proton exchange membrane fuel cells (PEMFCs), and suggests that minimization of particle movement can potentially slow down the coarsening of the particles, and the corresponding performance degradation.

Proton exchange membrane fuel cells (PEMFCs) represent an attractive energy conversion technology, which is also an environmentally-friendly alternative to the internal combustion engine used in automobiles[1,2]. In a PEMFC, hydrogen fuel is oxidized at the anode and oxygen is reduced at the cathode to produce water and electricity. While the standard reduction potential of $O_2$ to $H_2O$ is +1.23 V, the kinetics of the oxygen reduction reaction (ORR) are notoriously slow, giving rise to large overpotentials. As a result, much recent work has focused on the improvement of the cathode's electrocatalytic activity by synthesizing nanoparticles composed of platinum alloyed with a second element[3-6]. Among these, one of the most promising candidates is $Pt_3Co$, which has been shown both experimentally[7] and theoretically[8] to have a 2-4 fold enhancement in the ORR kinetics relative to pure Pt. However, the rapid degradation of the $Pt_3Co$ nanocatalyst results in unacceptably short life times of the fuel cell[1]. To date, the cost and degradation of cathode catalyst are two of the major barriers for the commercialization of PEMFCs. Therefore, it is of great importance to understand the degradation mechanisms of the nanocatalyst particles during electrochemical aging in a fuel cell cathode. A great deal of past efforts have focused on the *ex-situ* study of the nanocatalyst after electrochemical aging in a membrane-electrode-assembly (MEA, the electrochemical interface of a fuel cell) [9-11]. In these experiments, the MEA was sectioned for TEM analysis, making repeated observations on the same particles impossible. Consequently, particle trajectories were not available, making it difficult to unravel the concurrent coarsening effects of particle coalescence and Ostwald ripening[9-13]. In addition, particle-support interactions are also of great importance to the coarsening of the nanoparticles[14-16]. However, with 2-D images alone, one cannot determine how the catalyst particles are distributed on a 3-D fuel cell carbon support[17-20].

Herein, we present a 3-D tomographic method that allows us to track, in 3-D, the trajectories and morphological changes of individual Pt-Co nanocatalyst particles on a fuel cell carbon support, before and after electrochemical aging via potential sweeps. To achieve this, we carried out the electrochemical experiments directly on a TEM support grid[21-24]. In essence, a carbon-coated gold index grid was used as the working electrode in a three electrode electrochemical cell (Figure 1a). The fuel cell catalyst materials were deposited on the grid prior to the electrochemical experiments. The grid (Figure 1b) contained indexed windows, allowing us to perform electron tomography on exactly the same region of the catalyst material before (Figure 1c) and after (Figure 1d) aging, from which we were able to identify hundreds of nanoparticles in 3-D with precise one-to-one correspondence. This enabled us to study the coarsening evolution of the particles by tracking location changes on the support surface, observing changes in particle volume, and directly visualizing coarsening events. We found that, following extensive cycling of the Pt-Co catalysts under normal fuel cell operating conditions (potential scan from +0.6 V to +1.0 V for 30,000 cycles), the coarsening could be mainly attributed to coalescence of the nanoparticles. Furthermore, the observed coalescence and movement of the nanocatalyst particles was found not to be the result of carbon support degradation within our experimental resolution of 0.5-1.0 nm. In addition, in order to detect changes in the near surface composition of the catalysts, we also performed atomic-scale electron energy loss spectroscopic (EELS) imaging of the nanoparticles, from which we extracted the chemical composition of the near-surface volume of the particles after potential cycling.

In our study, a carbon-coated gold TEM grid with indexed windows (Figure 1b) was chosen as the working electrode. The indexed windows facilitated the determination of locations of the same catalyst materials before and after the electrochemical aging (Figure 1c) in a three-compartment electrochemical cell (Figure 1a). Gold grids were chosen because they are relatively inert over the potential range of interest. To demonstrate that the TEM grid is a reliable electrode, we characterized the grid from two aspects. First, Figure 1S (a) presents a comparison of the voltammetric profiles of a gold TEM grid and a gold disk electrode. Both profiles exhibited a Au-oxide reduction peak at approximately +1.2 V (vs. RHE)[*], which was characteristic of a gold surface. The double-layer region exhibited a capacitive response, which indicated that the surface of the grid was free of contamination.

---

[*] In the rest of the text, reported potentials are referenced to a RHE unless otherwise specified.

Second, the cross-talk between the catalyst material and the electrode should be minimal under the electrochemical aging conditions of interest, i.e. between +0.6 V to +1.0 V in our experiments. Figure 1S (b) shows that, below +1.4 V, interference from the gold TEM substrate was not significant, demonstrating the electrochemical reliability of the grid. In addition to the cycling setup, it is crucial to choose the appropriate electrochemical aging method. In the case of platinum, previous studies have suggested that potential cycling is more damaging than holding the potential at a given value[25, 26] (Figure 2S), and that a triangular wave form is more damaging than a square wave[27]. Therefore, we chose to cycle the potential with a triangular wave, so as to introduce more severe aging effects to the electrocatalyst nanoparticles.

The aging experiment was performed at a scan rate of 50 mV/s, between +0.6 V to +1.0 V (vs. RHE), and a complete cycle from +0.05 V to +1.2 V was carried out immediately after each thousand cycles to determine the electrochemical surface area (ECSA) of the catalyst. The choice of aging window of +0.6 V to +1.0 V and 30,000 cycles is based on the US Department of Energy (DOE) accelerated stress test protocols for PEMFCs[28]. The electrochemical aging results are shown in Figure 2a. Two key changes are highlighted by the blue and red dashed circles. First, the hydrogen adsorption peak at around +0.27 V in the anodic scan became sharper with aging. This sharpening effect was ascribed to the nanocatalyst surfaces becoming more Pt rich. The EELS spectrum analysis further confirmed such an assertion (vide-infra). Moreover, the hydrogen adsorption peaks at higher potential increased at the expense of those at lower potential. This is typically an indication of exposure of different facets after the electrochemical treatment[29]. The second feature to notice is the shift in the platinum oxide reduction peak at around +0.8 V to more positive values. To exclude the possibility that such a shift was due to the acid-leaching of cobalt, a set of control experiments was carried out on pure Pt nanocatalysts (i.e. no cobalt), as shown in Figure 3S. The results indicate that the positive shift of Pt oxide reduction peak could be observed both in $Pt_3Co/C$ and Pt/C. Consequently, we ascribe the positive shift of the Pt-O reduction peak not due to Co leaching. It was speculated that such shift could be caused by the increase in the size of the nanoparticles, as it has been previously reported that a growth in the size of the nanoparticles is often accompanied by a shift in the Pt-O reduction peak[30, 31]. In our study, accompanied by the size increase of some nanoparticles, the overall surface area decreased, as suggested by the ECSA measurements in Figure 2b. The ECSA was calculated by determining the charges associated with the adsorption and desorption of hydrogen between +0.05 V and +0.40 V—assuming 0.21 mC/cm$^2$ as the conversion factor for one monolayer of adsorbed hydrogen. Upon electrochemical aging, platinum was first oxidized to platinum oxide in the anodic scan (with an onset potential around +0.80 V), and the oxide was subsequently reduced back to platinum during the cathodic scan (with a peak potential at around +0.80 V). The dissolution-redeposition process results in a decrease in the ECSA[32]. At the end of the 30,000 cycles, the ECSA had dropped to around 87% of its initial value. This provides a direct indication of the overall changes in the surface area of the nanocatalysts.

The ORR polarization curves on the $Pt_3Co$/C nanocatalyst, before and after different extents of aging cycles, are shown in Figures 2c-d. The experiments were recorded on the same RDE electrode in an $O_2$-saturated 0.1 M $HClO_4$ solution at a rotation rate of 1600 rpm, after regular intervals of electrochemical aging. Metrics by which the activity of the nanocatalyst towards the ORR can be assessed include the potential at half maximum kinetic current ($E_{1/2}$), and the current at +0.85 V and +0.90 V. The last two can be used to evaluate the mass activity, by normalizing the kinetic current to the mass of the nanocatalyst on the electrode surface (Figure 3S for more detail). We found that the $E_{1/2}$ shifted in the negative direction by about 15 mV and that the mass activity at +0.85 V and +0.90 V decreased by 15 - 20% after 30,000 cycles of aging. Such a loss in ORR activity was consistent with the ECSA results. The degradation of the nanocatalyst was further confirmed by CO stripping (Figure 4S). The CO oxidation peak shifted to lower potentials, suggesting that the aging process resulted in an increase of the effective particle size. By comparing the integrated area of CO oxidation peak (Figure 4S), a 22% surface area loss was also observed, which was in agreement with the ECSA measurement.

Figures 5S-7S present bright-field TEM comparisons of the Pt-Co nanocatalyst materials. Owing to the labeled TEM tracing grid, it was possible to locate the same areas before and after electrochemical aging. It has been reported that carbon corrosion can play an important role in the degradation of Pt-based nanocatalysts in an MEA[15, 16], in which the potential at the cathode could go significantly above +1.0 V, especially under hydrogen starvation conditions-so the carbon support can be oxidized. However, that was not the case in our study. In the three-electrode system employed, control of the potential upper limit during potential cycling was accurate without ambiguity. As evident in the lower magnification images in Figures 7S, over the cycling range of +0.6 V to +1.0 V, no large-scale degradation of the carbon support was observed. From the higher resolution images used for the tomographic series (Figure 6s c-e), any changes to the support on a scale larger than 0.5 nm-1 nm could be excluded. However, we lack the 3D resolution to detect atomic-scale (i.e <0.5 nm) changes on the curved surfaces of the support. Therefore, we cannot rule out the possibility of atomic-scale changes of the support's surface at the particle-support anchoring point. In order to eliminate the possibility that the coarsening may be caused by other effects during the handle of TEM grid, a control experiment (Figure 2S) was performed in which the potential was held constantly at +1.2 V for 16 hours, after the electrochemical aging. Virtually no change could be observed from both TEM and ECSA measurements. Thus, the coarsening events were attributed to the potential cycling. In Figure 5S, the particles circled in yellow were suspected to have sintered during aging.

To further explore their chemical compositions, EELS mapping of those 17 spots, after electrochemical aging, was performed and the results are present in Figure 3 (The detailed methods of the EELS mapping can be found elsewhere[12, 33-35]). A Pt-rich shell was clearly observed, with an average shell thickness determined to be 0.85 ± 0.04 nm for single-core nanoparticles, and 0.93 ± 0.07 nm for multi-core nanoparticles. Both numbers correspond to a Pt shell thickness of 3-4 Pt monolayers (see Figure 8S for more details on determining the Pt shell thickness). This final shell thickness is less than the values in our previous study of the same starting material where the Pt shells were found to grow from 2-3 monolayers to 5-8 monolayers [12, 13]. This is also not unexpected as the aging conditions here were milder than our previous work[12, 13]: the upper potential limit was lowered to +1.0 V from +1.2 V and above, and the experiment was carried out at room temperature, not 80°C or higher.

EELS mapping is key for understanding how the Pt is redeposited, i.e., is the nanoparticle coarsening dominated by Ostwald ripening or coalescence? In Ostwald ripening, larger nanoparticles grow at the expense of smaller ones, with the small ones eventually being consumed. In such a scenario, a single-core particle should be expected[36, 37]. On the other hand in coalescence, two or more particles merge together to form a multi-core particle[36, 37]. Since only Pt should be able to redeposit over this potential range, the Pt-rich shell should be much thicker in Ostwald ripening when compared to coalescence. If Ostwald ripening played major role, then the shell thickness on multi-core particles should consequently be much larger than on single-core particles. However, as we determined in a statistical box plot (Figure 9S), no obvious difference in shell thickness between multi-core and single-core nanoparticles could be observed. Therefore, we conclude that the electrochemical aging in our study is mainly dominated by coalescence.

Taking advantage of the distinctive dissolution and redeposition properties of Pt and Co, we found that coalescence played a major role in the coarsening of the particles using atomic-scale chemical mapping. However, the trajectories of the coalesced particles were still missing, which made it difficult to understand the underlying driving force(s) for this particular coarsening mechanism. To overcome this limitation, the electron tomography [17-20] of a Pt-Co nanocatalyst region, obtained at the exactly same location both before and after electrochemical aging, is shown in Figure 4. The area was chosen to be representative of a typical particle distribution and carbon support morphology (Figures 6S-7S). By overlaying the reconstructions before and after the aging, we can directly visualize particle movement and coalescence (Figures 4a-b, and Movie 1S). It is immediately evident that most particles did not move after aging, which was confirmed by comparing the volume distribution and nearest neighbor distances (Figure 10S). The mean particle radius, around 3.2 nm, also did not change significantly—any changes that may have occurred were within the error of the measurement and represented no more than one Pt monolayer (Figure 12S). The preservation of particle positions greatly facilitated accurate alignment of the reconstructions in 3-D. The large set of fiduciary particles enabled an alignment fine enough to detect even small movements (~1-2 nm) on the carbon support. A more dramatic example of particle migration is visualized in Figure 4f, where a nanocatalyst particle 'falls' from the negative curvature (summit) into the positive curvature (valley) of the carbon support. We suspect this movement is likely driven by maximizing the contact area with the carbon support.

The electrochemical results (hydrogen adsorption regions in Figures 2a and 3S) suggest that the ratio of {100} to {111} facets increased during the cycling. Tracking particles one-by-one, before and after electrochemical aging, also reveals the coarsening mechanisms in the system. We found that the majority of the coarsening events were caused by particle coalescence, as illustrated in Figures 4c-e. The particles were seen to both migrate and coalesce with adjacent particles. This is in contrast to Ostwald ripening, where the particles dissolve and uniformly redeposit onto larger particles nearby. After coalescence, there is a further diminution of surface area that takes place during the dissolution-redeposition process when the connection between the coalesced particles is rounded off to minimize surface energy. This is best seen in Figure 4d, where two separate particles clearly merged uniformly. Such a result is in agreement with our previously proposed particle-particle-interaction between coalescence and Ostwald ripening[12, 13]. It is important to note that isosurfaces[#] can sometimes be misleading with ill-chosen thresholds. Therefore, in this study, all coalescence events were verified with the volumetric reconstruction data (Figure 11S, Movie 2S and 3S). In Figures 11S(d)-(e), a gap seen between two particles suggests they were separate before aging, while the disappearance of such a gap

---

# Isosurfaces represent points of constant value within a volume of space; it is a set of continuous function whose points group is in 3-D space (http://en.wikipedia.org/wiki/Isosurface)

confirms that the two particles were in contact with each other at the end of the cycles. This faithfully substantiates our visualization of nanoparticle coalescence in three-dimensions.

Although Ostwald ripening and coalescence might co-exist during electrochemical aging, it has been predicted that each of them would dominate at different potentials[27, 38]. At a higher cycling potential (i.e. in MEA with hydrogen starvation), the Pt dissolution-redeposition was so severe that larger nanoparticles consumed smaller ones quickly. On the other hand, at a lower cycling potential (i.e. our case), the migration of nanoparticles mainly took place to minimize the surface tension. One way to address the complex question would be to cycle the same nanoparticles to the potential higher than +1.4 V. However, under such conditions, our assumption that the gold TEM grid was stable within the electrochemical aging ranges would no longer be valid.

Moreover, it has been reported that one of the coarsening mechanisms in Pt/Pt-alloy nanoparticle systems is carbon corrosion[22, 39, 40]. However, no obvious carbon degradation at a scale larger than 0.5 nm-1.0 nm was observed in our study (Figures 6S and 7S), although we cannot fully rule out any atomic-scale changes that might occur on the surfaces of the carbon support. The possible explanations for the contrast with previous studies are as follow. In membrane electrode assembly (MEA) studies, the potential cycling window can be easily altered by hydrogen starvation and fuel cross-over. This can cause the potential at the cathode to be as high as +1.4V. However, in our well-defined three-electrode electrochemical system, the control of the cycling window (+0.6 V to +1.0 V) is accurate. Therefore, the kinetics of carbon corrosion in this cycling window could indeed be very slow.

In summary, we report for the first time a direct 3-D visualization of nanoparticle trajectories before and after electrochemical aging based on TEM grid indexing. The growth in the Pt shell thickness and observation of coalescence in 3-D could explain the decrease in electrochemically active surface area and loss of activity of Pt-Co nanocatalysts in fuel-cell cathodes. The simplicity and rigor of the three electrode system helps to rule out many factors that could take place in the MEA (such as hydrogen starvation), and therefore helps establish a direct correlation between ECSA and activity loss with the sintering behavior of nanocatalysts. The sintering processes were probed by atomic-scale chemical mapping and 3-D tomography. The majority of the coarsening events were determined to be coalescence, but not Ostwald ripening. These findings suggest that minimization of particle movement can potentially slow down the coarsening of particles during the electrochemical aging, and will be of great assistance in the future design of electrochemically durable fuel cell nanocatalysts and supports.

ACKNOWLEDGMENT

The $Pt_3Co/C$ and Pt/C were synthesized by Tanaka Kikinzoku Kogyo K. K. (TKK), Japan, and was generously provided by General Motor (Honeoye Falls, NY). We thank fruitful discussions with V. Liu, F. T. Wagner and M. F. Mathais on materials characterization. Y. Yu is grateful to J. L. Grazul, M. Thomas, H. Wang, and D. A. Finkelstein for helpful suggestion on TEM, EELS, electrochemistry, and RDE, respectively. Y. Yu thanks Z. Zou (Jimei University, China) for the assistance in art design and graphics. This work was supported by the Department of Energy though grant DE-FG02-87ER45298, by the Energy Materials Center at Cornell, an Energy Frontier Research Center funded by the U.S. Department of Energy, Office of Basic Energy Sciences under Award Number DE-SC0001086. This work made use of the electron microscopy facility of the Cornell Center for Materials Research (CCMR) with support from the National Science Foundation Materials Research Science and Engineering Centers (MRSEC) program (DMR 1120296).

**Supporting Information Available**

The electrochemical and TEM characterizations of nanocatalyst, before and after aging, are available on line at http://pubs.acs.org free of charge. Also included are three movies of nanoparticles on the carbon support.

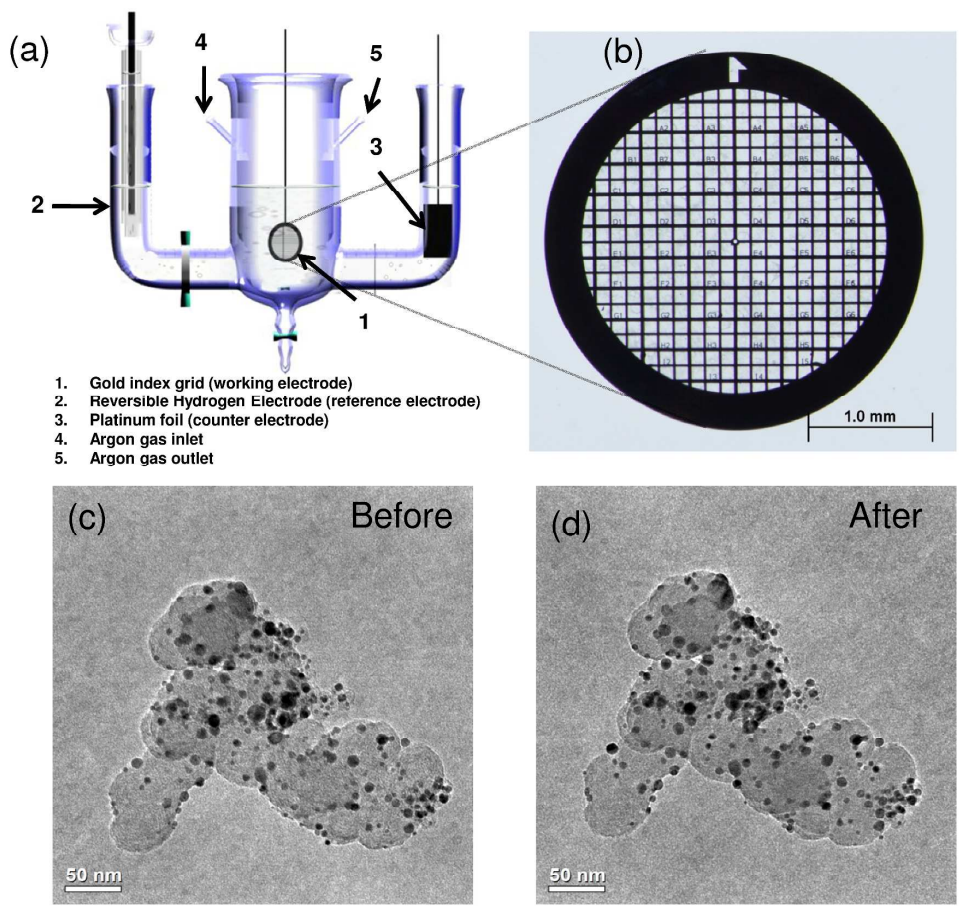

**Figure 1. (a)** Schematic of the three electrode electrochemical cell. **(b)** The carbon-coated gold finder grid. **(c)-(d)** Bright field TEM images of the same Pt-Co fuel cell material (metallic nanocatalyst loaded on 3-D carbon-black supports) identified before (c) and after (d) electrochemical aging.

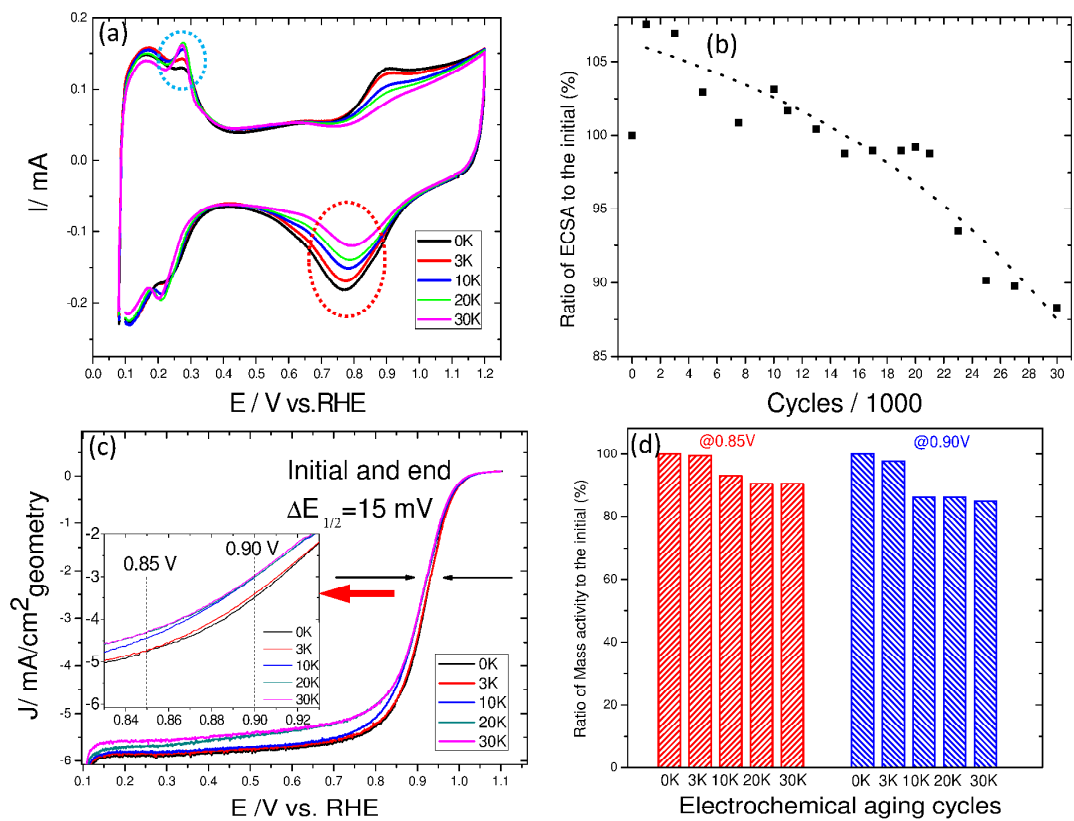

**Figure 2. (a)** Cyclic voltammogram (CV) of $Pt_3Co/C$ nanocatalyst, before electrochemical aging (0K), and after 3000 (3K, 'K' represents 1000), 10K, 20K and 30K cycles (between +0.6 V and +1.0 V) in 0.1M $HClO_4$ at a scan rate of 50 mV/s. The blue dashed-circle highlights the sharpening of the $H_2$ adsorption peak at ~ +0.27 V. The hydrogen adsorption peaks at higher potential increase at the expense of those at lower potential. It is an indication the {100} facets grow at the expense of {111} facets[29]. The red dashed-circle highlights the shift of the Pt-O reduction peak to more positive potentials. **(b)** Plot of electrochemical surface area (ECSA) as a fraction of its initial value, as a function of the number of cycles, for experiments performed on a gold grid. **(c)** RDE voltammograms of the oxygen reduction reaction (ORR) during the aging process, in $O_2$ saturated 0.1M $HClO_4$ solution, at a scan rate of 5 mV/s and rotating rate of 1600 rpm. Inset is the magnification from +0.83 V to +0.93V. The shift in the potential at half maximum kinetic current ($\Delta E_{1/2}$) was 15 mV. **(d)** Percentile of mass activity calculated at +0.85 V (red) and +0.90 V (blue) compared to its initial value, during the aging process.

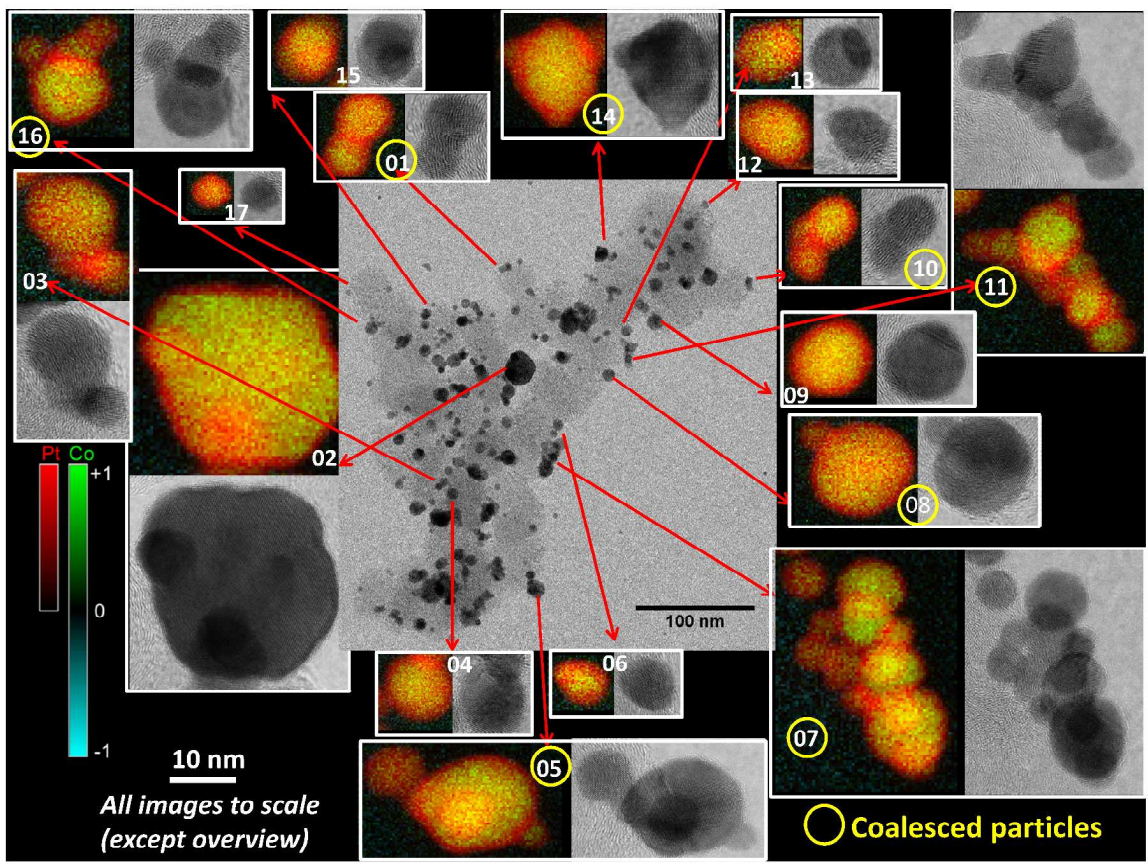

**Figure 3.** Pt$_3$Co/C nanocatalyst after electrochemical aging. The overall bright-field transmission electron microscopy (BF-TEM) image is shown in the center, with specific areas highlighted alongside Pt/Co chemical maps obtained by electron energy loss spectroscopy (EELS). The relative Pt concentration is shown in red and the Co concentration in green; yellow indicates a PtCo alloy. The arrow on the EELS map points out the position of the targeted particle from its carbon support. The yellow circle on the number indicates particles that were coalesced. All images (except the overview) are to the same scale for easy comparison.

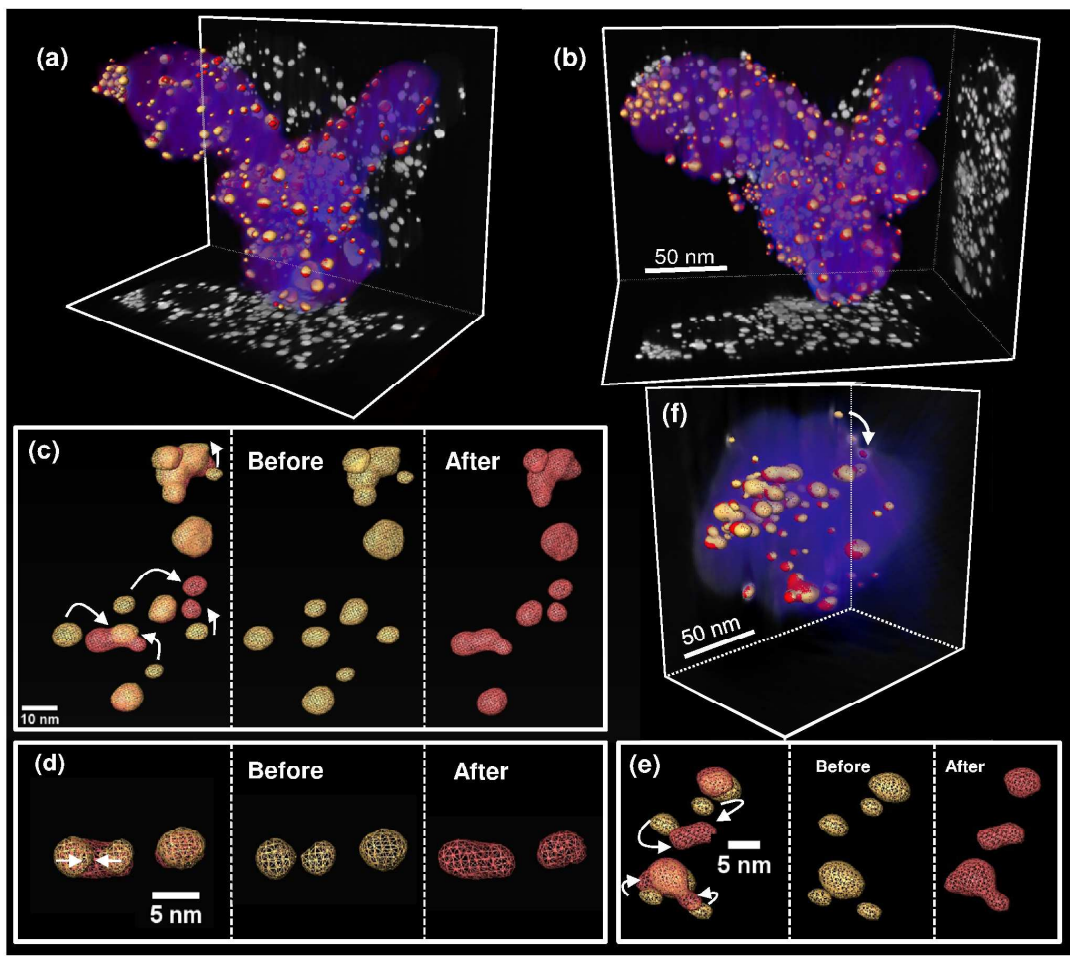

**Figure 4.** One-to-one correspondence of nanocatalyst particles before (*gold*) and after (*red*) electrochemical aging—same color scheme used in all figures. **(a)** 3-D reconstruction of nanocatalyst particles, on the carbon support (*violet*) with projected 2-D images shown at each side. **(b)** Alternate viewing angle. **(c,d,e)** Several instances of nanocatalyst particle coalescence and migration, with particle positions indicated by arrows. **(f)** One example of cropped volume from **(a,b)**, showing how one nanocatalyst particle moves into the positive curvature (valley) from the negative curvature (summit) of the catalyst support. The arrow points out the trajectory of the particle movement. Violet shading is the carbon support.

**TOC Graphic**

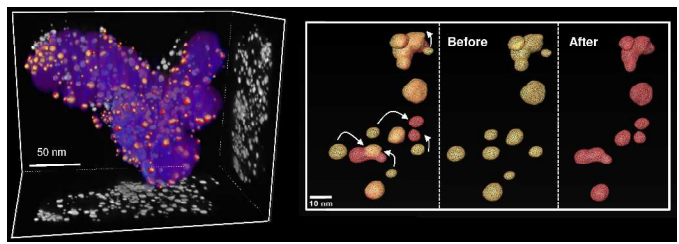

# Supporting information

### 3-D Tracking and Visualization of Hundreds of Pt-Co Fuel Cell Nanocatalysts During Electrochemical Aging

*Yingchao Yu, Huolin L. Xin, Robert M. Hovden, Deli Wang, Eric D. Rus, Julia Mundy,*

*David A. Muller, Héctor D. Abruña*

EXPERIMENT

1. **Materials.**

    The two catalyst samples employed in the experiments, $Pt_3Co$ and Pt, were synthesized by Tanaka Kikinzoku Kogyo K. K. (TKK), Japan. Both samples were supported on Vulcan® XC-72 carbon black support, with Pt loadings of approximately 30 and 46 weight percent, respectively.

2. **Electrochemical testing**.

    The electrochemical experiments were performed using a PAR 273 potentiostat. All electrochemical experiments were carried out in 0.1 M $HClO_4$ (70 weight percent, from Sigma-Aldrich) electrolyte at room temperature, unless otherwise indicated. All solutions were made with ultra-pure Millipore water (18.2 MΩ/cm). To prepare the working electrode, the $Pt_3Co$/C or Pt/C nanocatalyst were first dispersed in an isopropanol/water solution (3:1 volume ratio). The concentration of the particles was controlled to be around 5mg $_{catalyst}$/ mL. Nafion (0.05 weight percent) was used to enhance the adhesion of the catalysts to the working electrode. After sonicating for 30 minutes, about 5.0 μL of solution was pipetted onto a labeled 2.5 mm diameter TEM grid (purchased from Electron Microscopy Sciences), as shown in Figure 1b. In a control experiment, the same catalyst suspension was deposited onto a 3 mm diameter glassy carbon

working electrode, to form a uniformly distributed thin layer. Such Nafion-catalyst-carbon inks adhered strongly to the electrode and served as the working electrode. The reference electrode was a reversible hydrogen electrode (RHE), against which all potentials are reported. A large area Pt wire was used as the counter electrode. Prior to electrochemical aging, the solution was purged with argon for 30 minutes, to remove any oxygen in the system. The setup in Figure 1a was air tight and argon was purged periodically, to minimize any interference from air during the experiment. The complete electrochemical setup is illustrated in Figure 1a. Oxygen reduction was carried out on a 3 mm diameter glassy carbon Rotating Disk Electrode (RDE), which was modified and treated in the same way. Oxygen was purged for at least one hour to saturate the solution. The polarization curves were obtained by sweeping the potential from +0.1 V to +1.1 V at a scan rate of 5 mV/s, with a rotation rate of 1600 rpm. Carbon monoxide stripping was carried out by first bubbling the solution with CO for 30 minutes and then with nitrogen for 30 minutes, while the potential was held at +0.05 V. Two cycles at 50mV/s between +0.05 V and +1.2 V were performed immediately after the CO purging.

3. **Electron Microscopy**

The tomographic tilt series were collected on a Schottky-field-emission-gun Tecnai F20 scanning transmission electron microscope (STEM) operated at 200keV. A high-angle annular dark field (HAADF) detector provided an incoherent projection image of the specimen with a signal intensity proportional to the amount of material and its atomic number, which is also known as Z-contrast. Images were collected at every two degrees of rotation, under the range of -70 to +72 degrees. The alignment and reconstruction of all tomography images was performed with in-house software written in Matlab. The visualization of the reconstructed data set was done in Avizo 6.3. The EELS data was acquired on a $5^{th}$-order aberration-corrected scanning transmission electron microscope (Nion UltraSTEM)operated at 100 kV, with a convergence angle $\alpha_{max}$ = ~30 mrad. The upper limit of beam dose was calculated to be $5.8 \times 10^6$ e⁻/nm² for a complete tomography experiment and $2.0 \times 10^3$ e⁻/(nm² s) for each second of electron beam exposure. Within such upper electron beam dose limit, no migration of nanoparticles could be seen (Figure 2S). To calculate the upper limit of electron beam dose, the following equation was used.

$$Unit\ Dose \left(\frac{e^-}{nm^2 \cdot sec}\right) = \frac{I \times N_A}{F \times A}$$

$$Total\ dose \left(\frac{e^-}{nm^2}\right) = \frac{I \times N_A}{F \times A} \times T$$

$$T(sec) = (t_1 \times 1024^2 + t_2) \times n$$

| Symbol | Notation | Value |
|---|---|---|
| I | Beam current | 10 pA |
| $N_A$ | Avogadro's number | $6.02 \times 10^{23}$ /mol |
| F | Faraday constant | $9.65 \times 10^4$ C/mol |
| A | Total area (Figure 12S) | $3.0 \times 10^4$ nm² |
| $t_1$ | Dwell time | 32 μs |
| $t_2$ | Time needed to focus image | 7s |
| n | Number of images | 71 |
| T | Total time of tomography | 2840 s |

4. **EELS analysis**

The Co $L_{2,3}$-edge was extracted using a power law background subtraction method. The integration window was set to catch both $L_2$ and $L_3$ edge of Co. Since the Pt $N_3$-edge has delayed onsite, a multivariate curve resolution (MCR) method was used to refine the spectra, in order to get better signal to noise ration. The detail about using MCR to analyze Pt $N_3$-edge could be found in our previous papers[1].

**Movie1S**. 3-D reconstructed movie of Pt$_3$Co/C, before and after aging. The nanoparticles are colored as golden (before) and red (after) while the carbon support is in violet.

**Movie 2S.** The real images while moving the oblique slice, before aging. See Figure 11S for the method. At the maximum contrast intensity, a gap could be seen between the two particles.

**Movie 3S.** The real images while moving the oblique slice, after aging. The gap seen in Movie 2S disappears.

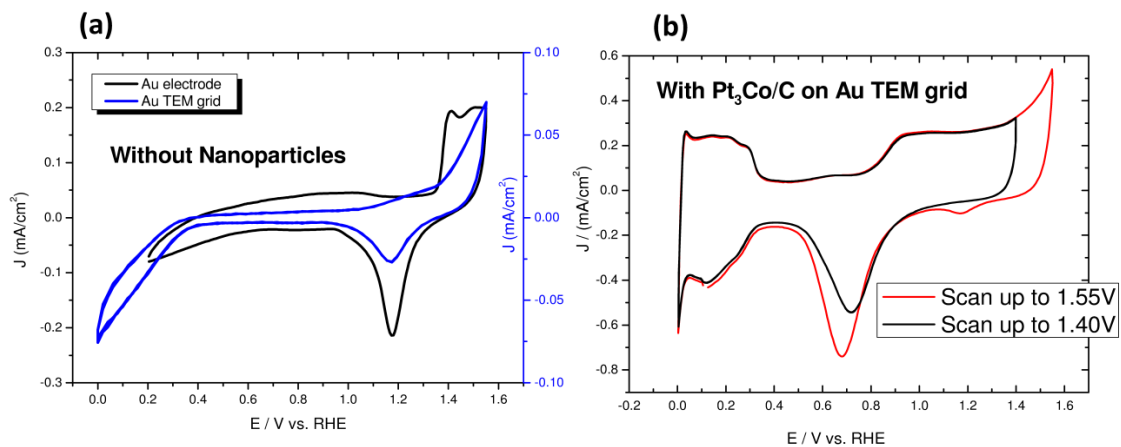

**Figure 1S**. **(a)** Cyclic voltammogram (CV) of a bulk gold electrode (black) and a gold TEM grid (blue), in 0.1M $H_2SO_4$, at a scan rate of 50 mV/s. Different scales of current density have been applied for easier comparison. The reduction peak at around +1.2V indicates the presence of gold on the surface, and the double-layer-only response suggests the electrode is free of contamination. **(b)** CV of $Pt_3Co/C$ on the gold TEM grid, in 0.1M $H_2SO_4$, at a scan rate of 50 mV/s, with different upper limits of +1.40V (black) and +1.55V (red).

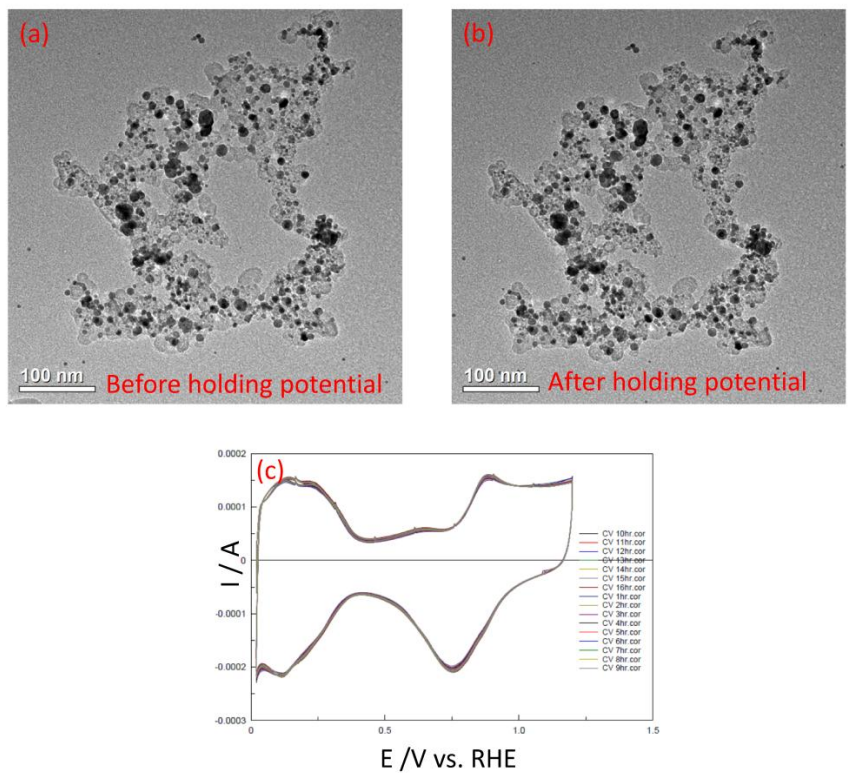

**Figure 2S.** The effect of holding a constant potential at +1.2V vs. RHE for 16 hours on the $Pt_3Co/C$ nanoparticles. **(a)-(b)** TEM image of one area before and after applying the constant potential, respectively; Almost no nanoparticle migration could be observed during the experiment; The image in Figure 2S(b) was taken after being exposed to electron beam exposure for about two hours. **(c)** CVs of the corresponding sample, in 0.1M HClO4, at a scan rate of 50 mV/s, performed at the interval of each hour, up to 16 hours. The overlap of CVs at different time indicates almost no drop on the electrochemical active surface (ECSA).

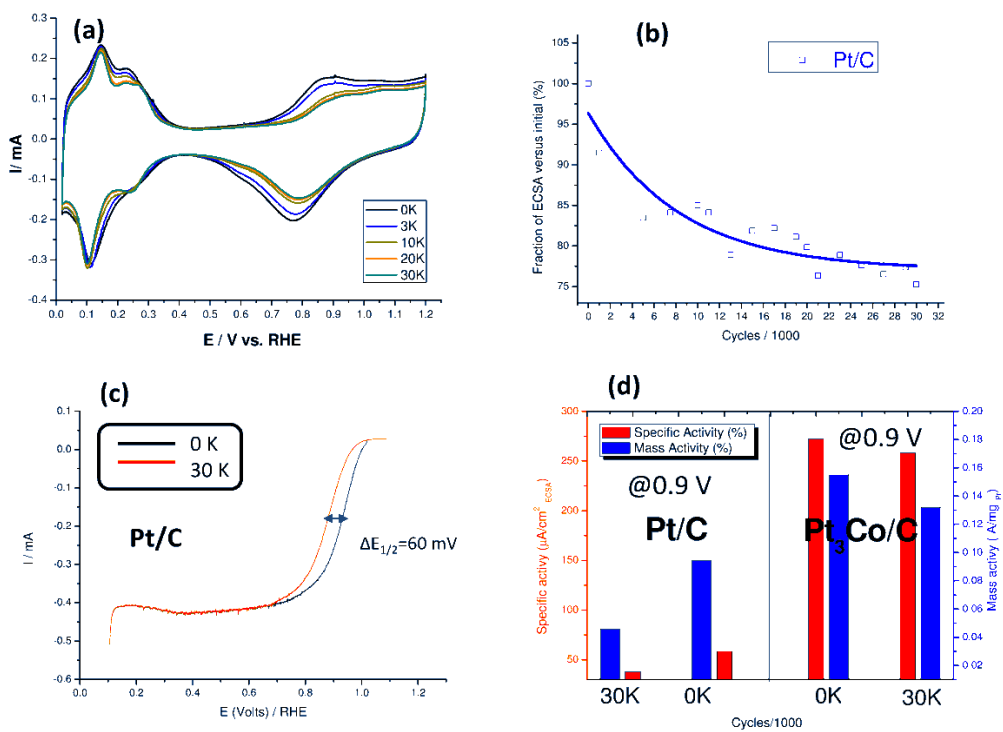

**Figure 3S.** **(a)** CVs of Pt/C in 0.1M $HClO_4$, after 0, 3K, 10k, 20k, 30k cycles of electrochemical aging, at a scan rate of 50 mV/s. **(b)** The percentile of electrochemical surface area (ECSA) relative to its initial value, as a function of aging cycles, for Pt/C. **(c)** RDE voltammograms of the oxygen reduction reaction (ORR) for Pt/C, in an $O_2$ saturated solution of 0.1M HClO4, at a scan rate of 5 mV/s and rotating rate of 1600 rpm, before (black) and after (red) aging. The change in $\Delta E_{1/2}$ is about 60 mV. **(d)** Comparison of specific activity (red) and mass activity (blue) for Pt/C (left) and $Pt_3Co/C$ (right). Both of $Pt_3Co/C$ and Pt/C had a drop in activity showing similar trends in degradation.

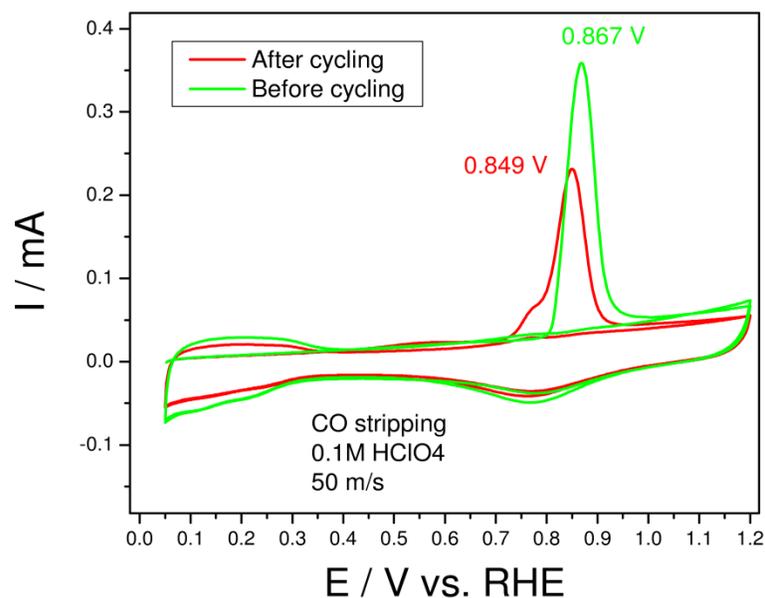

**Figure 4S.** CO stripping of Pt$_3$Co/C before and after aging, in 0.1M HClO$_4$ solution. The experiment was carried out by purging CO for half of hour, while the potential was held at +0.05V. Nitrogen was then purged for another 30 minutes to remove CO from the bulk solution. Two cycles between +0.05V and +1.2V were performed immediately after the nitrogen purging, at a scan rate of 50mV/s. The second cycle showed elimination of CO, indicating the absence of CO on the electrode. The negative shift of the oxidation peak is 18 mV, an indication of growth of the effective nanoparticle size as the adsorption energy of CO decreased. The integrated area of CO oxidation between +0.7V and +1.0V exhibited a 22% surface area loss, which is consistent with the hydrogen adsorption measurement.

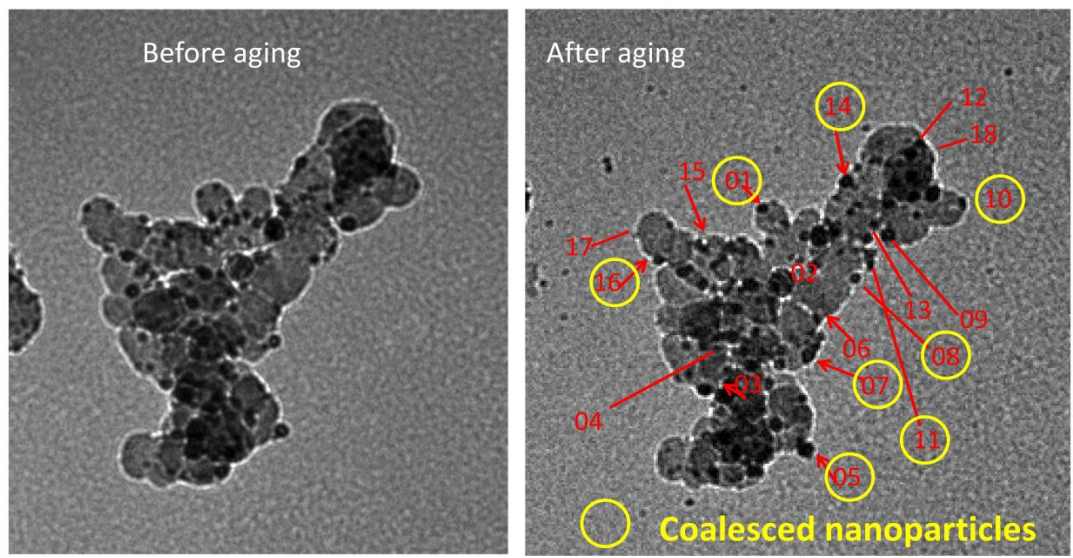

**Figure 5S.** Bright field TEM comparison of one region of $Pt_3Co/C$ nanocatalyst before and after aging. The possible coalesced particles are highlighted by yellow circles. The same number sequence is also used in Figure 3.

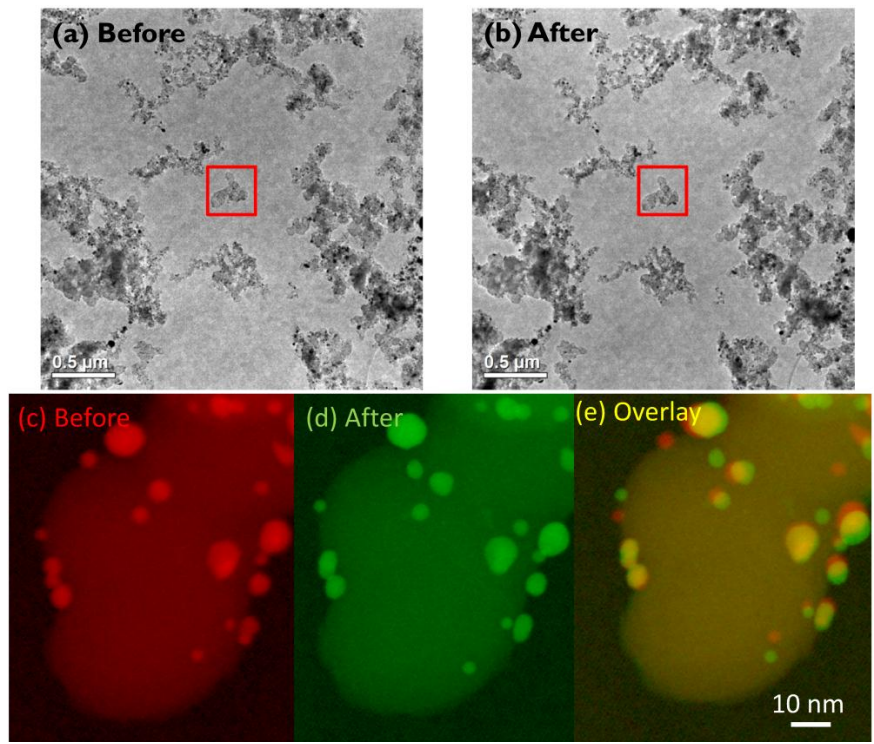

**Figure 6S. (a)-(b)** Bright field TEM overview of Pt$_3$Co/C before and after electrochemical aging, respectively. Note the tomography (shown in Figure 4a-b) area is right in the middle. **(c)-(e)** HAADF-STEM images of the upper left portion of the highlighted regions in (a)-(b). By comparison, no obvious nano-scale carbon degradation can be observed within the resolution of measurement (0.5 nm-1.0 nm). Evidences of nanoparticles coalescence could also be found by comparing (c) and (d) in the overlay (e) – yellow regions are unchanged, but red and green regions show the before and after location of particles.

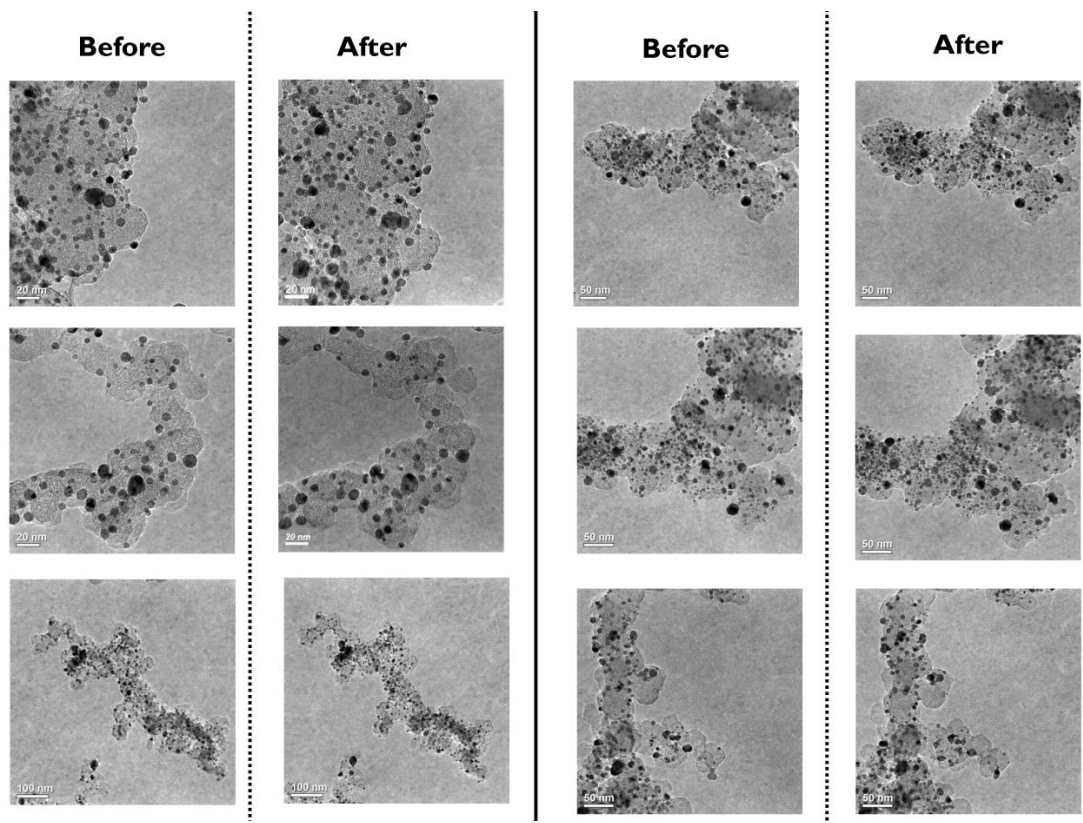

**Figure 7S.** Bright field TEM of six carbon support areas, before and after electrochemical aging. The results are shown as two separate columns. The sampling covers the major shapes shown in Figure 6S, and indicates the choice of area where tomography experiment was performed, was a representative of the majority of nanoparticles dispersed on the catalyst support.

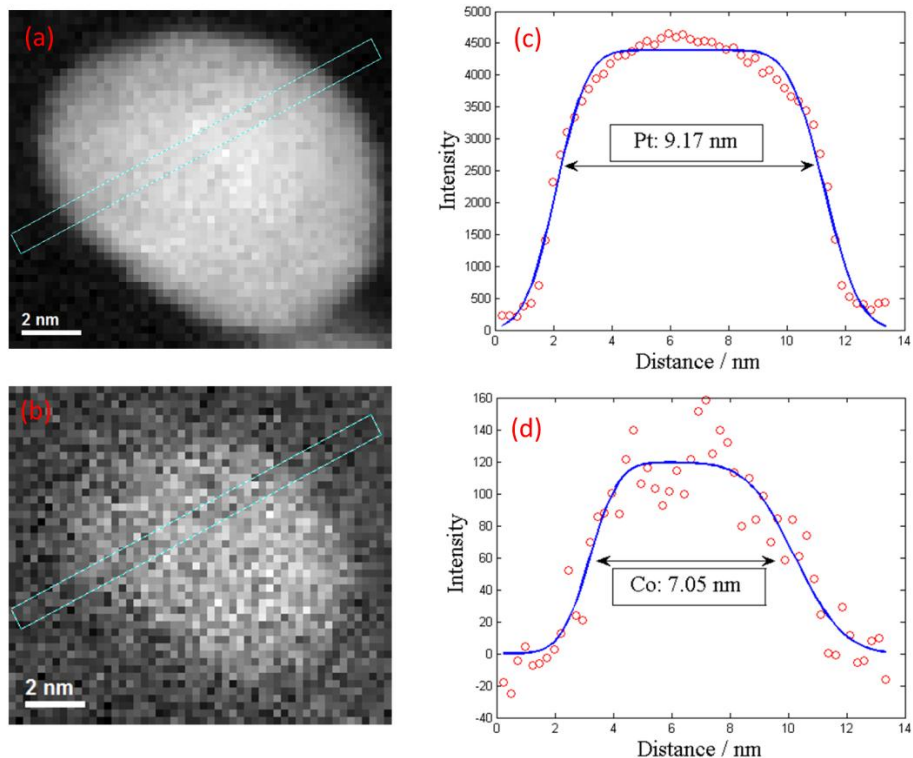

**Figure 8S**. One example of using EELS to analyze the Pt shell thickness. **(a)-(b)**, The EELS mapping of signals from Co and Pt, respectively; **(c)-(d)**, the corresponding best fit of the Co and Pt signals with two error functions, $b\dfrac{Erf[a_1(x-x_1)] + Erfc[a_2(x-x_2)] - 1}{2}$, with fitting parameters $b$, $a_1$, $a_2$, $x_1$ and $x_2$. The full width at half maximum (FWHM) can be extrapolated from the two error functions, to calculate the length of each element (denoted as d). The Pt shell thickness was calculated as $d_{shell} = \dfrac{d_{Pt} - d_{Co}}{2}$. The example above shows how to calculate the shell thickness based on this method. For instance, if $d_{Pt}$=9.17nm and $d_{co}$= 7.05nm, then the Pt shell thickness is obtained as $d_{shell}$= (9.17 nm -7.05 nm)/2= 1.06 nm.

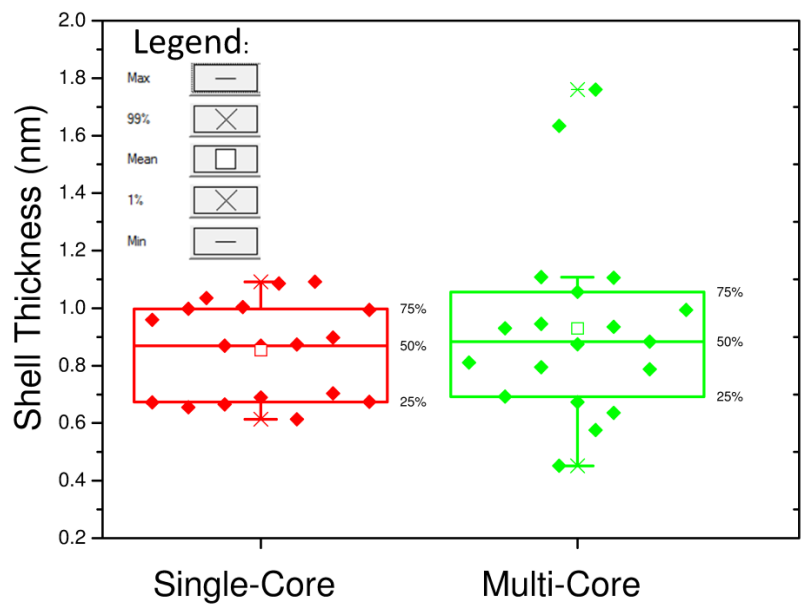

**Figure 9S**. The box plot of statistically analyzed Pt shell thickness, for single-core (red) and multi-core (green) nanoparticles. The maximum and minimum, 99% and 1%, mean values of the data have been displayed as lines, crosses, and squares, respectively. Single-core and multi-core nanoparticles show no obvious difference in shell thickness.

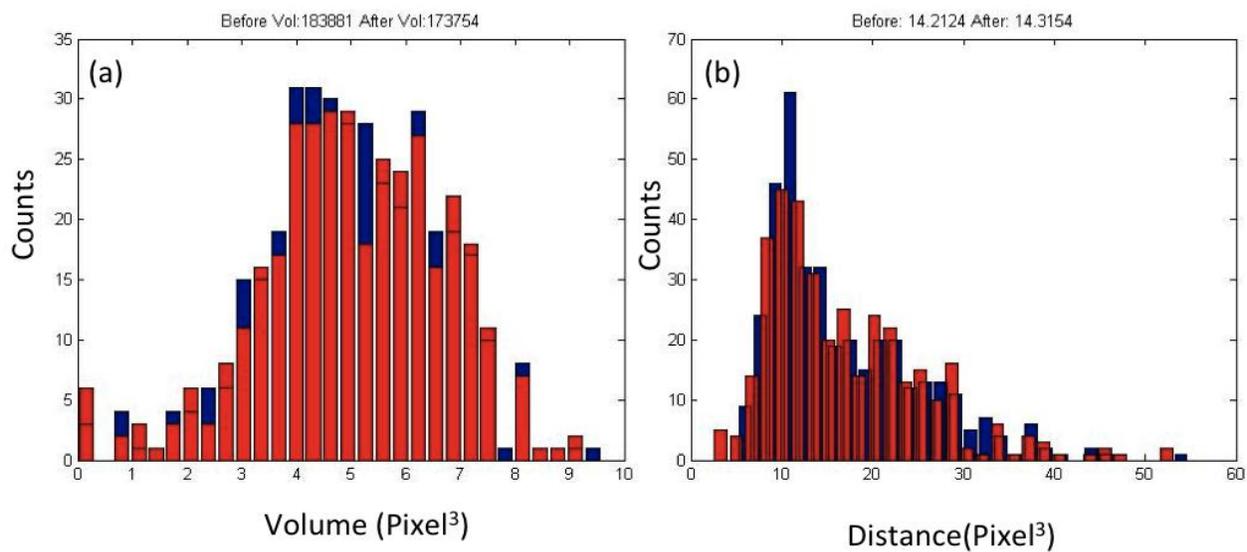

**Figure 10S (a).** Volume distribution of Pt$_3$Co/C before (red) and after (blue) electrochemical aging. **(b)** Nearest neighbor distance before (red) and after (blue) the aging. Little difference is observed between the two sets of data. The analysis results are tabulated in Figure 12S.

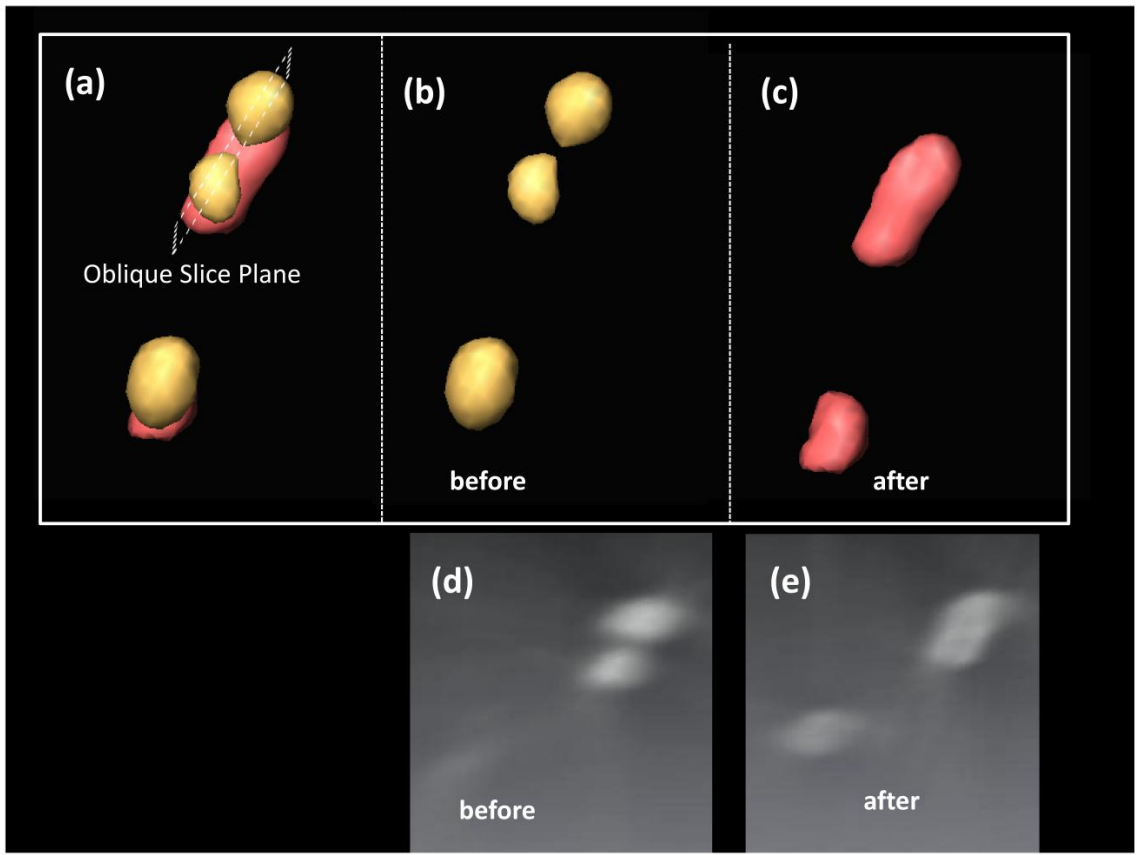

**Figure 11S. (a)-(c)** Schematics of how to use an oblique slice to check the threshold of particles isosurfaces. A rectangular shape plane is moved perpendicular to the zone axis of nanoparticles, and the real dark field images are viewed simultaneously. The before aging is colored as gold while the after aging is colored red. **(d)-(e)** Real images at the maximum intensity of moving the oblique slice. A clear gap is seen before aging, indicating the separation of these two particles, and the disappearance of such a gap suggests coalescence.

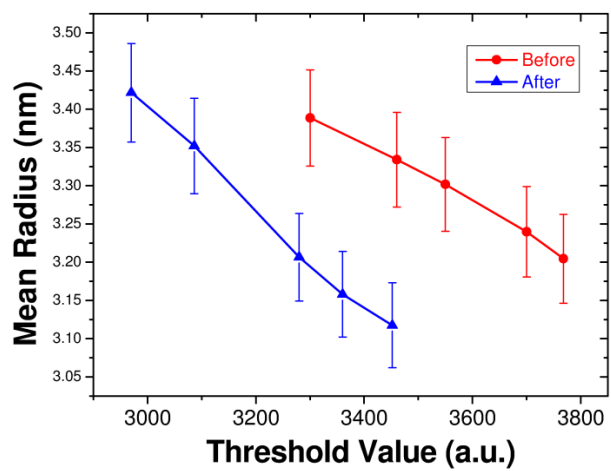

|                                      | **Before**  | **After**   | Ratio of after versus before |
|--------------------------------------|-------------|-------------|------------------------------|
| Number of NPs                        | 382         | 377         | 0.99                         |
| Total Volume (nm$^3$)                | 57321       | 53578       | 0.93                         |
| Mean Radius (nm)                     | 3.29±0.03   | 3.23±0.06   | 0.98                         |
| Total Surface Area(nm$^2$)           | 31995       | 30822       | 0.96                         |
| Nearest Neighbor Distance (nm)       | 9.93        | 10.02       | 1.01                         |

**Figure 12S**. Influence of different threshold values on the mean radius of the nanoparticles, before and after electrochemical aging. The final threshold value was chosen based on the average number of five different trials. The error bar corresponds to two standard error of the mean, at each data point, with a 95% statistical confidence interval. The maximum change on nanoparticles' mean radius is one atomic monolayer, upon choosing different threshold values. Inserted table: the statistical analysis of Figures 10S and 11S.